\documentclass[aps,nofootinbib,pra,amsmath,amssymb,twocolumn]{revtex4-2}
\pdfoutput=1
\usepackage{amssymb}
\usepackage{color}
\usepackage{graphicx}
\usepackage{epsfig,amssymb,amsmath}
\usepackage{chngcntr}
\usepackage{enumitem}
\usepackage[normalem]{ulem}

\graphicspath{ {./images/} } 
\usepackage[caption=false]{subfig} 
\usepackage{dsfont}

\usepackage{color}
\newcounter{bar}

\newcounter{num}

\def\>{\rangle}
\def\<{\langle}
\def\comment#1{}
\def\labell#1{\label{#1}}
\def\togli#1{}

\begin{document}
\title{Tight Entropic Uncertainty Relations} \author{Alberto
  Riccardi}\email{Corresponding Author: alberto.riccardi13@gmail.com.}
  \author{Lorenzo Maccone$^1$} \affiliation{\vbox{$1.$~Dip.~Fisica and INFN
    Sez.\ Pavia, University of Pavia, Via Bassi 6, I-27100 Pavia,
    Italy}}
\begin{abstract}
  Entropic uncertainty relations $H(A)+H(B)\geqslant \gamma$ give a
  nonzero lower bound $\gamma$ to the sum of the Shannon entropies $H$
  of the outcome probabilities of incompatible observables $A$ and
  $B$. They are better than the variance-based uncertainty relations
  because they only depend on the Born statistics of the outcomes and
  not on the outcomes themselves, and because bounds $\gamma$
  typically are state independent. Here we provide a state-independent
  lower bound $\gamma_s$ that is better than the textbook
  Maassen-Uffink bound and, in the limit of the parameter $s\to 2$,
  becomes asymptotically tight for all $A,B$. The bound can be
  extended to Renyi entropies.

{\bf  After the submission of this paper, we found out that the main
  results presented here were already on page 6 of
  \cite{rene}.}
\end{abstract}
\maketitle

Entropic uncertainty relations \cite{review1,review2,review3} take the
form $H(A)+H(B)\geqslant\gamma$ with $H(X)=-\sum_jp(x_j)\log_2p(x_j)$
the Shannon entropy of the outcome statistics $p(x_j)$ of the
observable $X$. For quantum observables, it takes the Born rule
expression $p(x_j)=|\<x_j|\psi\>|^2$ with $|x_j\>$ the eigenstates of
$X$ and $|\psi\>$ the system state. The entropic relations are well
suited to capture the true essence of incompatibility between
observables $A$ and $B$ because the bound $\gamma$ typically depends
only on the eigenstates $\{|a_i\>\}$ and $\{|b_j\>\}$ of $A$ and $B$
and not on their eigenvalues. This is notable because the Born rule
probabilities depend only on the eigenstates and the observables
eigenvalues are only used to track the measurement outcomes. Moreover,
the lower bound $\gamma$ is typically independent on the state of the
system, in contrast to the variance-based Heisenberg-Robertson
\cite{heisenberg,robertson} ones
$\Delta^2A\Delta^2B\geqslant\tfrac14|\<\psi|[A,B]|\psi\>|^2$, where
$\Delta^2X=\<X^2\>-\<X\>^2$ and the lower bound depends explicitly on
the system state $|\psi\>$.  Entropic uncertainty relations were first
introduced by Hirschmann \cite{hirschmann}, and Bialynicki-Birula and
Miycielski \cite{bialynicki}. They were then refined by Deutsch
\cite{deutsch} and finally Maassen and Uffink \cite{muffink} derived the ones that
are typically used today: $H(A)+H(B)\geqslant-2\log_2c$, with
$c=\max_{ij}|\<a_i|b_j\>|$. This bound only uses the maximum overlap
between the eigenstates: a very limited information on $A,B$. More
recent results \cite{colespiani,karol} use also the second largest
overlap. In this paper we provide a bound that uses the {\em full}
unitary matrix of overlaps $U_{ji}=\<b_j|a_i\>$:
\begin{align}
  H(A)+H(B)\geqslant\frac{-2s}{2-s}\log_2\|\<b_j|a_i\>\|_{s\to s'}
\labell{bound}\;,
\end{align}
for all $s\in(1,2)$ and $1/s+1/{s'}=1$. Here
$\|U\|_{p\to q}=\sup_{\psi_i}\tfrac{\|U\psi\|_{q}}{\|\psi\|_p}$ is the
$p\to q$ operator norm of $U$ with
$\|x\|_\ell=(\sum_j|x_j|^\ell)^{1/\ell}$. The bound is state
independent, and nontrivial for all incompatible observables $A,B$
that do not share any eigenstates. In the limit $s\to2^-$, the bound
is tight: the right-hand-side tends to the minimum of the sum of
entropies. Other values of $s\in(1,2)$ give worse bounds, that are all
stronger than the Maassen-Uffink \cite{muffink} one.  The
right-hand-side of \eqref{bound} can be numerically calculated
efficiently using the nonlinear power iteration method (NPIM)
\cite{boyd,hing}, reviewed in App.~\ref{s:npim}. Thanks to this, it is
possible to have a good estimate of the minimum sum of entropies for
observables in large-dimensional Hilbert spaces, which would otherwise
be practically impossible (Fig.~\ref{f:dim}). For qubits a partially
closed form exists \cite{zozor}, see App.~\ref{s:qbit}.

\begin{figure}[hbt]
\begin{center}
\epsfxsize=.55\hsize\leavevmode\epsffile{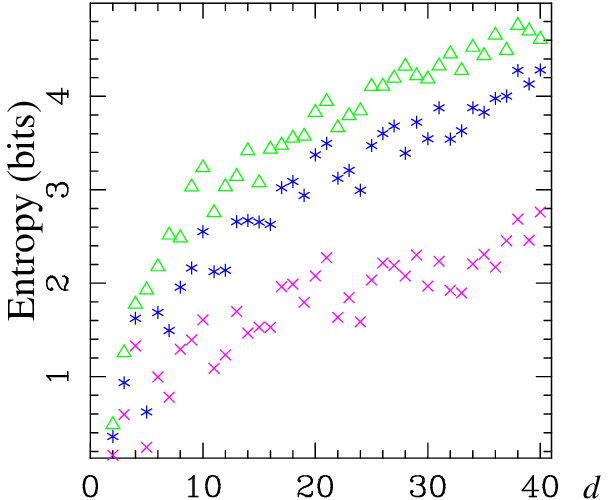}
\end{center}
\vspace{-.5cm}
\caption{The bound \eqref{bound} (blue stars) for $s\to2$ estimates
  the minimum sum of entropies also for large dimension $d$. The
  purple crosses are the Maassen-Uffink bound, whereas the green
  triangles is the sum of entropies when one of the states is the
  lowest sum-of-entropies eigenstate of one observable. Here $s=1.95$
  and for each $d$ a new unitary $U$ is chosen randomly according to
  the Haar measure \cite{karol1,karol2}. The bound is estimated using
  NPIM with $5\times 10^4$ random seed states for each $d$.}
\labell{f:dim}\end{figure}

The bound \eqref{bound} can also be extended to Renyi entropies
\cite{renyi,review1,review2,renyireview} defined by 
$H_\alpha(X)\equiv\tfrac1{1-\alpha}\log_2\sum_i|x_i|^{2\alpha}$, as
\begin{align}
H_{s/2}(A)+H_{s'/2}(B)\geqslant\frac{-2s}{2-s}\log_2\|\<b_j|a_i\>\|_{s\to s'}
\labell{renyi}\;,
\end{align}
for all $s\in(1,2)$, and $1/s+1/s'=1$. These are tight for all $s$! In
particular, in the limit $s\to2$ it gives back the best among the
bounds \eqref{bound}, when both Renyi entropies in \eqref{renyi} tend
to the Shannon one.

\section{Proof of the bound} The bound \eqref{bound} is an application
of the Riesz–Thorin interpolation theorem \cite{dunford}
(incidentally, also the Maassen-Uffink bound \cite{muffink} is an
application of the Riesz–Thorin theorem): Consider a linear operator
$T$ that is bounded both according to the $p_0\to q_0$ and the
$p_1\to q_1$ norms: $\|{T}\|_{p_i\to q_i}\leqslant M_i$, $i=0,1$. Then
the following upper ``interpolating'' bound to the
$p_\theta\to q_\theta$ norm holds
\begin{align}
  \|{T}\|_{p_\theta\to q_\theta}\leqslant (M_0)^{1-\theta}(M_1)^\theta,\qquad\mbox{ with }\labell{rt}\\
  \tfrac1{p_\theta}=\tfrac{1-\theta}{p_0}+\tfrac\theta{p_1},\ \tfrac1{q_\theta}=\tfrac{1-\theta}{q_0}+\tfrac\theta{q_1},
\nonumber
\end{align}
for all $\theta\in[0,1]$.

Consider the operator $U_{ji}=\<b_j|a_i\>$, which is unitary because
it transforms the basis $\{|b_j\>\}$ of eigenstates of $B$ into the
basis $\{|a_i\>\}$ of eigenstates of $A$. Its $\ell^2$-norm is
$\|U\|_2=1$, because $\|a_i\|_2=\|b_i\|_2=1$ due to the Born rule
probability normalization, with $a_i\equiv\<a_i|\psi\>$,
$b_j\equiv\<b_j|\psi\>=\sum_iU_{ji}a_i$ [namely, $|x_i|^2=p(x_i)$, $x=a,b$]. Use the Riesz–Thorin inequality for
$q_1=p_1=2$ to obtain
\begin{align}
  \frac{\|b\|_{q_\theta}}{\|a\|_{p_\theta}}=\frac{\|Ua\|_{q_\theta}}{\|a\|_{p_\theta}}\leqslant\sup_a\frac{\|Ua\|_{q_\theta}}{\|a\|_{p_\theta}}=
  \|U\|_{p_\theta\to q_\theta}\nonumber\\
  \leqslant\|U\|^{1-\theta}_{p_0\to q_0}\|U\|^\theta_{p_1\to q_1}=\|U\|^{1-\theta}_{p_0\to q_0}
\labell{ineq}\;,
\end{align}
which, taking the $\log$, becomes
\begin{align}
\log_2\|b\|_{q_\theta}-\log_2\|a\|_{p_\theta}\leqslant(1-\theta)\log_2\|U\|_{p_0\to
  q_0}
\labell{logg}\;.
\end{align}
Choose $p_0=s$ and $q_0=s'$ and divide both members by
$1-\theta>0$. Take the limit $\theta\to 1^-$ using the definition of
the derivative: since $\log_2\|b\|_{q_1}=\log_2\|a\|_{p_1}=0$, we
find \begin{eqnarray} \lim_{\theta\to
    1^-}-\tfrac{\log_2\|a\|_{p_\theta}}{1-\theta}=\tfrac\partial{\partial\theta}\log_2\|a\|_{p_\theta}\Big|_{\theta=1}
  =-\tfrac{2-s}{2s}H(A) ,\ 
\labell{lim}\;
\end{eqnarray}
and similarly for $\log_2\|b\|_{q_\theta}$ whose derivative gives
$-\tfrac{2-s}{2s}H(B)$. This statement and the last equality in
\eqref{lim} are proven in App.~\ref{s:eqlim}. The bound \eqref{bound}
is finally found by dividing both members of \eqref{logg} by $-\tfrac{2-s}{2s}<0$,
which flips its inequality.  $\square$
\section{Bound tightness}
We now prove that, in the limit $s\to 2$, the bound \eqref{bound} is tight, namely that
\begin{align}
\lim_{s\to 2^-}\frac{-2s}{2-s}\log_2\|U\|_{s\to s'}=\inf_{|\psi\>}[H(A)+H(B)]
\labell{tight}\;.
\end{align}

Writing explicitly $\|U\|_{s\to s'}=\sup_\psi\|U\psi\|_{s'}/\|\psi\|_s$ in the left-hand-term and considering that the $\log$ is a growing function, while the term $-2s/(2-s)<0$, the sup becomes an inf:
\begin{align}
  \lim_{s\to 2^-}\tfrac{-2s}{2-s}\log_2\|U\|_{s\to s'}=\inf_\psi  \lim_{s\to 2^-}\tfrac{-2s}{2-s}\log_2
  \tfrac{\|U\psi\|_{s'}}{\|\psi\|_s}
\labell{a1}\;.
\end{align}
With a change of variables $x\equiv 2-s>0$, and remembering that $s'=s/(s-1)$, the right-hand-side becomes
\begin{align}
  \inf_\psi\lim_{x\to 0^+}\tfrac{-2(1-x)}x\log_2{\textstyle\sum_{i}|\sum_j}U_{ij}\psi_j|^{\tfrac{2-x}{1-x}}\nonumber\\
  +\tfrac2x\log_2{\textstyle\sum_i}|\psi_i|^{2-x}
\labell{a2}\;.
\end{align}
Using the trick of App.~\ref{s:eqlim}, these two limits can be turned into
derivatives. Indeed, the second term is equal to
\begin{align}
2\tfrac\partial{\partial
  x}\log_2\sum_i|\psi_i|^{2-x}\Big|_{x=0}=-\sum_i|\psi_i|^2\log_2|\psi_i|^2
\labell{a3}\;,
\end{align}
which is equal to $H(A)$ by choosing $\psi_i=\<a_i|\psi\>$. Similarly,
the first term in \eqref{a2} is equal to
$-\sum_{i}|\sum_jU_{ij}\psi_j|^2\log_2|\sum_jU_{ij}\psi_j|^2$, which is
equal to $H(B)$ for $\psi_j=\<a_j|\psi\>$, since $b_i=\sum_jU_{ij}a_j$. This proves \eqref{tight}, since the
minimization over $|\psi\>$ is equivalent to the minimization over
$\psi_j$. $\square$

The tightness of the bound for $s\to2^-$ is explored numerically in
Fig.~\ref{f:comparison} both using Montecarlo techniques and NPIM.

\begin{figure}[hbt]
\begin{center}
  \epsfxsize=1.\hsize\leavevmode\epsffile{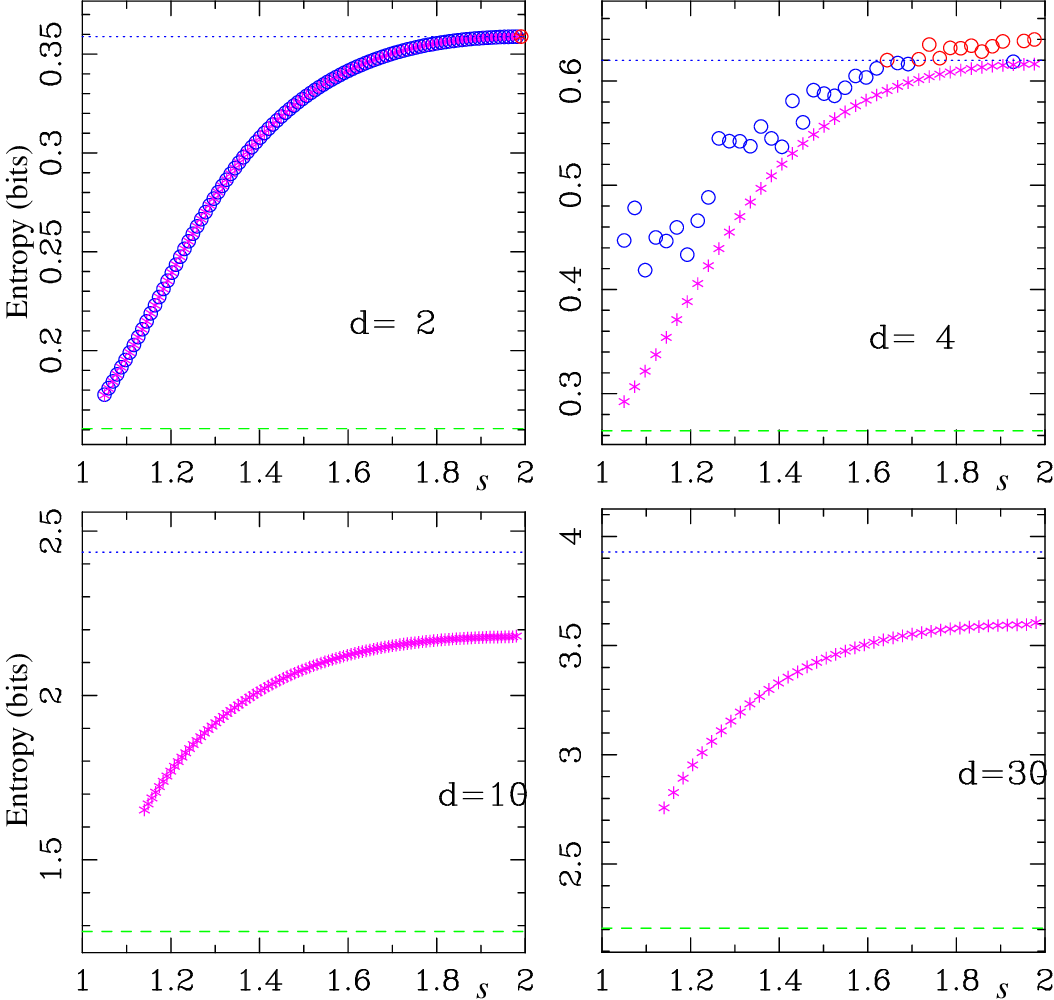}
\end{center}
\vspace{-.5cm}
\caption{Upper: comparison between Montecarlo and NPIM as a function
  of $s$ for dimension $d=2$ (left) and $d=4$ (right). The calculation
  of the bound \eqref{bound} $s\to s'$ norm can be done by estimation
  of the sup over randomly generated states (circles) or using the
  NPIM method \cite{hing,dunford} (stars). The dashed line is the
  Maassen-Uffink bound, the dotted line is the minimum sum-of-entropy
  of all the considered states. Here $10^8$ random states were used
  for the Montecarlo, while only $10^3$ random seed states for
  NPIM. Many of the Montecarlo states fall above the lower bound (red
  circles) showing that Montecarlo is inadequate already for
  $d=4$. Lower: the bound as a function of $s$ calculated with NPIM
  for $d=10,30$, with $10^4$ seed states. Again the dashed line is the
  Maassen-Uffink bound and the dotted line is an {\em upper bound} to
  the minimum sum-of-entropy calculated by minimizing it over $10^8$
  random states (but no state with lower sum-of-entropy than the
  eigenstate of one of the observable was found!). In all cases, a
  unitary $U$ was chosen randomly \cite{karol1,karol2}.}

\labell{f:comparison}\end{figure}
\section{Comparison with Maassen-Uffink}\labell{s:muffink}
We now show that the bound \eqref{bound} is stronger than the
Maassen-Uffink one (the latter is tight if the observables $A,B$ have
mutually unbiased eigenbases). Take $c=\max_{ji}|U_{ji}|$, then for all $a_i$ and all $j$
\begin{align}
  |{\textstyle\sum_i}U_{ji}a_i|\leqslant{\textstyle\sum_i}|U_{ji}|\,|a_i|\leqslant
  c{\textstyle\sum_i}|a_i|=c\|a\|_1
\labell{al}\;,
\end{align}
which (maximizing over $j$) implies $\|U\|_{1\to\infty}\leqslant
c$. Moreover, unitarity implies $\|U\|_{2\to2}=1$. So we can employ
Riesz–Thorin \eqref{rt} with $p_0=1$, $q_0=\infty$, $p_1=q_1=2$, which
gives $\|U\|_{p_\theta\to q_\theta}\leqslant c^{1-\theta}$. Choose
$s=p_\theta$ and $s'=s/(s-1)=q_\theta$. Taking the log of both
members, the Riesz–Thorin inequality becomes \begin{align}
\log_2\|U\|_{s\to s'}\leqslant \tfrac{2-s}s \log_2c
\labell{rt2}\;,
\end{align}
since $1-\theta=\tfrac{2-s}s$. Multiplying both members of \eqref{rt2}
by $\tfrac{-2s}{2-s}<0$ we find that our inequality is stronger than
Maassen-Uffink's: \begin{align}
  H(A)+H(B)\geqslant\tfrac{-2s}{2-s}\log_2\|U\|_{s\to
    s'}\geqslant-2\log_2c \labell{muf}\;.
\end{align}
\section{Renyi entropies}\labell{s:renyi}
We now prove bound \eqref{renyi}. Again, use the Riesz–Thorin bound
with $p_0=s$, $q_0=s'$, $p_1=q_1=2$ and choose $\theta=0$ in
\eqref{logg}, namely
$\log_2\|b\|_{s'}-\log_2\|a\|_s\leqslant\log\|U\|_{s\to s'}$. This is
equivalent to \eqref{renyi}, because for $s'=s/(s-1)$ the definition
of the Renyi entropy implies \begin{align}
  -\log_2\|a\|_s=&-\tfrac{2-s}{2s}H_{s/2}(A)\;,\nonumber\\
  \log_2\|b\|_{s'}=&-\tfrac{2-s}{2s}H_{{s'}/2}(B) \labell{kk}\;.
\end{align}
The tightness of the bounds \eqref{renyi} follows from the definition
of $\|U\|_{s\to s'}$. Indeed, \begin{align}
  \tfrac{-2s}{2-s}\log_2\sup_{\psi_i}\tfrac{\|U\psi\|_{s'}}{\|\psi\|_s}=
  \inf_{\psi_i}\tfrac{-2s}{2-s}\log_2\tfrac{\|U\psi\|_{s'}}{\|\psi\|_s}
  \nonumber\\
  =\inf_{|\psi\>}[H_{s/2}(A)+H_{s'/2}(B)]
\labell{tightr}\;,
\end{align}
where we used \eqref{kk} in the last equality and the fact that
minimizing over $\psi_i$ is equivalent to minimizing over $|\psi\>$ as
discussed above.  $\square$

In particular, this implies that the Maassen-Uffink lower bound
$-2\log_2c$ of \eqref{muf} is a tight lower bound for the Renyi
entropy uncertainty relations with $s=1$ and $s'=\infty$, or with
$s=\infty,s'=1$. Indeed, $\|U\|_{1\to\infty}=c$, since \eqref{al}
implies $\|U\|_{1\to\infty}\leqslant c$, and one can also prove the
``$\geqslant$'' by considering a vector $e$ that is composed by all
zeros except for a one in correspondence to the index $i$ for which
the $\max_{ji}|U_{ji}|=c$ is realized, so that
$\|Ue\|_\infty\geqslant c$, while $\|e\|_1=1$.

\section{Conclusions}
In conclusion, we have provided the lower bound \eqref{bound} to the
sum of entropies of the quantum statistics of the outcomes of two
arbitrary observables $A$ and $B$. We have shown that it is tight in
general, in contrast to all known bounds
\cite{muffink,deutsch,colespiani,karol}, and that the right-hand-side
of the inequality \eqref{bound} can be numerically evaluated using
NPIM. Finally, we have shown that the bound can also be extended to
Renyi entropies as Eq.~\eqref{renyi}, which are all tight. We have
discussed pure states here, but the bounds apply also to mixed states,
since mixed states trivially have larger entropies.

Why the bound? The inequality \eqref{renyi}, which comprises the best
among the ones of \eqref{bound}, is tight, so there must be a deep
reason why its lower bound is what it is. Indeed, we have shown that
it comes from the Riesz-Thorin inequality with all its free parameters
fixed: $p_0=s$ and $q_0=s'$ are determined by the choice of the Renyi
entropies we are considering, while $p_1=2$ is determined by the fact
that the Born rule uses {\em square} moduli of the amplitudes and not
a different power \cite{sorkin} which is an, experimentally determined
\cite{urbasi}, postulate of quantum mechanics. Instead, $q_1=2$
because we are considering observables $A,B$, which are associated to
eigenbases (another postulate of quantum mechanics), so a unitary $U$
is required to move from one basis to the next\comment{Ovviamente,
  fare cadere $p_1=q_1=2$ e' il primo step per fare una relazione
  indet per POVMs!}. Finally, the fact that $\tfrac1s+\tfrac1{s'}=1$
is only a request for elegance: we want a bound that only involves
sums of entropies, if we settle for sums of entropies weighted by
appropriate coefficients, one can drop this last request and
immediately obtain an amended version of \eqref{renyi}.

{\em Acknowledgements}. This paper is dedicated to Matilde who was born during its
development. L.M. acknowledges support from the National Research
Centre for HPC, Big Data and Quantum Computing, PNRR MUR Project
CN0000013-ICSC.

\newpage
\appendix
\section{NONLINEAR POWER ITERATION METHOD}\labell{s:npim}
In this appendix we review the nonlinear power iteration method, also
referred to in the literature as ``power method for $\ell^p$ norms''
or ``nonlinear power method'' \cite{boyd,hing}. It was used to
calculate the plots in Figs.~\ref{f:dim} and \ref{f:comparison}. The
algorithm's goal is to find the maximum value of $\|Ux\|_{s'}$ subject
to the constraint $\|x\|_s = 1$, namely, the operator norm
$\|U\|_{s \to s'}$, with $s\in(1,2)$ and $s'=s/(s-1)>2$.

{\begin{center}{\bf NPIM Algorithm:}\end{center}}\vspace{-.3cm}
\begin{enumerate}[wide, labelwidth=!, labelindent=5pt]
\item\label{seedsi} Generate a random complex vector (seed) $u^{(k)}\in{\mathbb C}^d$ with $k=0$.
\item\label{iterationsi} Normalize it to create the vector $v^{(k)}=u^{(k)}/\|u^{(k)}\|_p$.
\item Apply the operator $U$: $w^{(k)}_j=\sum_iU_{ji}v^{(k)}_i$.
\item Calculate the norm for the $k$-th iteration as $\|w^{(k)}\|_{s'}$.
\item Rescale it as $x^{(k)}_j=w^{(k)}_j\:|w^{(k)}_j|^{s'-2}$, where $s'-2>0$, since $s'>2$.
\item Apply the adjoint to go back: $y_i=\sum_{j}U^*_{ji}x_j$.
\item Rescale it again as $z^{(k)}_j=y^{(k)}_j\:|y^{(k)}_j|^{s'-2}$.
\item Iterate from step \ref{iterationsi}, choosing $v^{(k+1)}=z^{(k)}$,
  and stop when the iteration converges:
  $\left|\,\|w^{(k+1)}\|_{s'}-\|w^{(k)}\|_{s'}\right|<\epsilon$, with $\epsilon$ some
  (small) accuracy parameter.
\item Repeat the whole procedure $N_{seeds}$ times from step
  \ref{seedsi} and select the {largest} $L$ of the obtained norms
  $\|w^{(k)}\|_{s'}$. This is needed because the algorithm converges
  to a {\em local} maximum and one needs to repeat it multiple times
  with $N_{seeds}$ random seeds to find the global maximum. 
\end{enumerate}
The final quantity $L$ is the estimated norm $\|U\|_{s\to s'}$. A good
check to see whether $N_{seeds}$ is sufficiently large is to calculate
the norm for different $s$ and check whether there are large
fluctuations for adjacent values of $s$.

\section{Qubits}\labell{s:qbit}
In this appendix we give a partially closed formula of the bound for
qubits, Eq.~\eqref{boundqbit} below, which was studied in
\cite{zozor}, and give an analysis for this case.

Without loss of generality, by appropriately choosing the $z$-axis of the Bloch sphere to coincide with the eigenstates of the operator $B$ and the $x$ axis to cancel all the phases in the eigenstates of $A$, we can write the unitary transition matrix $U_{ji}=\<b_j|a_i\>$ as a real rotation matrix
\begin{align}
  U=\begin{pmatrix}\cos\varphi&\sin\varphi\\
  -\sin\varphi&\cos\varphi\end{pmatrix},
  \qquad\varphi\in\bigl(0,\tfrac{\pi}{4}\bigr].
\labell{unitqub}\;,
\end{align}
where $\cos\varphi\geqslant0$, $\sin\varphi\geqslant0$. We must
maximize the norm of $U$ when evaluated over an arbitrary 2-$d$
vector $e$ with $\|e\|_s=1$, namely (apart from irrelevant phases):
\begin{align}
e = \begin{pmatrix} \cos^{2/s}(\alpha) \\ e^{i\gamma} \sin^{2/s}(\alpha) \end{pmatrix}
\labell{qbit}\;.
\end{align}
Taking $y_j = \sum_iU_{ji}e_i$, we find
\begin{align}
|y_1|^2 = \cos^2(\varphi) \cos^{4/s}(\alpha) + \sin^2(\varphi)
\sin^{4/s}(\alpha) \nonumber\\+ 2
\cos(\varphi)\sin(\varphi)\cos^{2/s}(\alpha)\sin^{2/s}(\alpha)
\cos(\gamma),\nonumber\\
|y_2|^2 = \sin^2(\varphi) \cos^{4/s}(\alpha) + \cos^2(\varphi)
\sin^{4/s}(\alpha) \nonumber\\- 2
\cos(\varphi)\sin(\varphi)\cos^{2/s}(\alpha)\sin^{2/s}(\alpha)
\cos(\gamma)
\labell{nor}\;,
\end{align}
which we can write more compactly as $|y_1|^2 = M_1 - I \cos(\gamma)$,
$|y_2|^2 = M_2 + I \cos(\gamma)$. The $s'$-norm of $y$ is then
\begin{align}
  (\|y\|_{s'})^{s'}= \left( M_1 - I \cos(\gamma) \right)^{s'/2} + \left( M_2 + I \cos(\gamma) \right)^{s'/2}
  \labell{normy}\;,
\end{align}
where $s' \ge 2$ so the above norm is a convex function with respect
to $\cos(\gamma)$, so the maximum of \eqref{normy} must lie on the
boundary, namely for $\cos\gamma=\pm 1$. In turn, this implies that
$\gamma=0,\pm\pi$ and we can limit ourselves to real vectors in
\eqref{qbit}, namely \begin{align}
x=\binom {\cos\alpha}{\sin\alpha}\ \Rightarrow\ Ux=\binom{\cos(\varphi-\alpha)}{\sin(\alpha-\varphi)}
\labell{qstate}\;.
\end{align}
Then, from the definition of $\|U\|_{s\to s'}$: \begin{align}
  \|U\|_{s\to s'}=\max_{\alpha\in[0,\pi/2]}\frac{(|\cos(\varphi-\alpha)|^{s'}+|\sin(\varphi-\alpha)|^{s'})^{1/s'}}
  {(\cos^s\alpha+\sin^s\alpha)^{1/s}}
\labell{maxma}\;,
\end{align}
and the bound \eqref{bound} becomes
\begin{align}
  H(A)+H(B)\geqslant\min_{\alpha\in[0,\pi/2]}B_s(\alpha),
  \nonumber\\\mbox{ with }
B_s(\alpha)\equiv\tfrac{-2s}{2-s}\log_2\|U\|_{s\to s/(s-1)}.
\labell{bound11}\;
\end{align}
As a function of $s$, the best of these bounds is obtained for
$s\to 2$. To calculate this limit, it is better to reparametrize using
$u_\alpha\equiv\cos^2\alpha$ and
$w_\alpha\equiv\cos^2(\alpha-\varphi)$, so that
\begin{align}
  B_s(\alpha)=\tfrac{-2s}{2-s}[&\tfrac1{s'}\log_2(w_\alpha^{s'/2}+(1-w_\alpha)^{s'/2})
               \nonumber\\&-\tfrac1s\log_2(u_\alpha^{s/2}+(1-u_\alpha)^{s/2})]
\labell{aks}\;,
\end{align}
where the function $f_r(p)\equiv\tfrac1r\log_2(p^{r/2}+(1-p)^{r/2})$
appears twice, with $p\in[0,1]$. To calculate the $s\to 2$ limit of
\eqref{aks}, we take the Taylor expansion of $f_r(p)$ around $r=2$,
where $f_2(p)=0$ and
\begin{align}
  \tfrac\partial{\partial r}f_r(p)\Big|_{r=2}=
  \tfrac1{r^2}\log_2(p^{r/2}+(1-p)^{r/2})+\nonumber\\
  \tfrac{p^{r/2}\log_2p+(1-p)^{r/2}\log_2(1-p)}{2r(p^{r/2}+(1-p)^{r/2})}\Big|_{r=2}=
  -\tfrac14 h(p)
\labell{de}\;,
\end{align}
where $h(p)= -p\log_2p-(1-p)\log_2(1-p)$ is the
binary entropy. So,
$f_{2\pm\epsilon}(p)\simeq
f_2(p)\mp\tfrac{\epsilon}4h(p)+O(\epsilon^2)$. Then choose $s=2-\epsilon$, so that
$s'=\tfrac{s}{s-1}=\tfrac{2-\epsilon}{1-\epsilon}\tfrac{1+\epsilon}{1+\epsilon}\simeq
2+\epsilon+O(\epsilon^2)$. Then,
\begin{align}
\lim_{s\to 2^-}B_s(\alpha)=h(\cos^2\alpha)+h(\cos^2(\alpha-\varphi))\equiv F(\alpha)
\labell{asda}\;,
\end{align}
because $\tfrac{-2s}{2-s}\simeq-\tfrac{4}\epsilon$ for
$\epsilon\simeq0$. Finally, \eqref{asda} must be minimized over
$\alpha\in[0,\pi/2]$, see \eqref{bound11}. To find the minimum, consider 
\begin{align}
  \tfrac\partial{\partial\alpha}F=&
  -
  2\log_2|\tan\alpha|\sin(2\alpha)\nonumber\\&-2\log_2|\tan(\alpha-\varphi)|\sin(2(\alpha-\varphi))
  \labell{derqu}\;,
\end{align}
where we used $\tfrac\partial{\partial p}h(p)=\log_2\tfrac{1-p}p$. It
is null for $\alpha=\varphi/2$. This value of $\alpha$ is not always a minimum, as can
be seen by looking at the second derivative, calculated in
$\varphi/2$:
\begin{align}
  \tfrac{\partial^2}{\partial\alpha^2}F\big|_{\alpha=\varphi/2}=
  -\tfrac8{\ln 2}[1+\cos\varphi\ln\tan(\varphi/2)]
\labell{der2qu}\;.
\end{align}
The second derivative can be obtained by defining
$g(\alpha)\equiv\log_2|\tan\alpha|\sin(2\alpha)$ so that
$\tfrac\partial{\partial\alpha}F(\alpha)=-2g(\alpha)-2g(\alpha-\varphi)$,
and then deriving $g$. It is strictly negative for all $\varphi$
smaller than a critical angle $\varphi_c\simeq 0.586$ radians.  Thus,
for $\alpha=\varphi/2$, $F(\alpha)=2h(\cos^2\tfrac\varphi2)$ is the
minimum sum of entropies for $\varphi\leqslant\varphi_c$. Instead, for
$\varphi>\varphi_c$ there is no closed form of the bound, which must
be calculated numerically. In essence, the bound is
\begin{align}
H(A)+H(B)\geqslant
\min[2h(\cos^2\tfrac\varphi2),\min_{\alpha\in[0,\pi/2]} F(\alpha)]
\;,\labell{boundqbit}
\end{align}
with $h$ the binary entropy, where the minimum is attained at the
first term for $\varphi\leqslant\varphi_c\simeq 0.586$, and at the
second otherwise. Eq.~\eqref{boundqbit} is validated in
Fig.~\ref{f:qbit} with a Montecarlo simulation that, for each
$\varphi$, calculates the minimum sum-of-entropy by minimizing it over
a large set of randomly generated qubit states.

\begin{figure}[hbt]
\begin{center}
  \epsfxsize=1.\hsize\leavevmode\epsffile{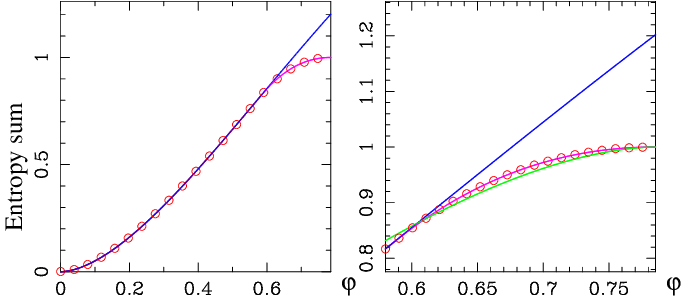}
\end{center}
\vspace{-.5cm}
\caption{Left: Depiction of the qubit lower bound \eqref{boundqbit} as
  a function of $\varphi$. The blue curve is the first term of the
  min, which is a minimum for all
  $\varphi\leqslant\varphi_c\simeq 0.586$, at which point the magenta
  curve (second term of the min) takes over. The latter is calculated
  by numerically finding the minimum of $F(\alpha)$, a single valued
  function. The red circles are a Montecarlo validation by minimizing
  the sum of entropy over $10^5$ randomly generated qubit
  states. Right: blow-up of the previous graph around $\pi/4$. The
  (green) curve $1-2h(\cos^2\tfrac{\pi/4-\varphi}2)$ is added to show
  that such curve is not a good bound, namely
  $2h(\cos^2\tfrac\varphi2)$ is not symmetric.}
\labell{f:qbit}\end{figure}

\section{Proof of Eq.~\eqref{lim}}\labell{s:eqlim}
In this section we prove the last equality of Eq.~\eqref{lim},
$-\tfrac\partial{\partial\theta}\log_2\|a\|_{p_\theta}\Big|_{\theta=1}=-\frac{2-s}{2s}H(A)$. We
also show that
$+\tfrac\partial{\partial\theta}\log_2\|b\|_{q_\theta}\Big|_{\theta=1}=-\frac{2-s}{2s}H(B)$.

Since $p_{1}=2$, then $\log_2\|a\|_{p_1}=0$, so, using the definition of the derivative,
\begin{align}
-\lim_{\theta\to 1^-}\frac{\log_2\|a\|_{p_\theta}}{1-\theta}=\frac{\partial}{\partial\theta}\log_2\|a\|_{p_\theta}\Big|_{\theta=1}
\labell{ab}\;.
\end{align}
Moreover, since $1/p_\theta=(1-\theta)/p_0+\theta/p_1$, with $p_0=s$ and $p_1=2$, then $\tfrac{\partial p_\theta}{\partial\theta}|_{\theta=1}=2\tfrac{2-s}{s}$, so
\begin{align}
  \frac\partial{\partial\theta}\log_2\|a\|_{p_\theta}\Big|_{\theta=1}
  =\frac{\partial p_\theta}{\partial\theta}\frac{\partial\log_2\|a\|_{p_\theta}}{\partial p_\theta}\Big|_{\theta=1}\nonumber\\
  =2\frac{2-s}{s}
  \frac\partial{\partial p}\frac1{p}\log_2\sum_i|a_i|^{p}\Big|_{p=2}\nonumber\\
  =2\frac{2-s}{2s}\sum_i|a_i|^2\log_2|a_i|=-\frac{2-s}{2s}H(A)
  \labell{aa}\;.
\end{align}

A similar procedure applies to the term with $\|b\|_{q_\theta}$, the
only difference being that $q_0=s'=s/(s-1)$. So, using the same
procedure of Eq.~\eqref{aa}, swapping $p_\theta$ with $q_\theta$, we
find
$\tfrac\partial{\partial\theta}\log_2\|b\|_{q_\theta}\Big|_{\theta=1}=-\tfrac{2-s}{2s}H(B)$.

[The choice $p_0=s$ and $q_0=s'=s/(s-1)$ is necessary in order to
obtain the same coefficients with opposite signs in front of $H(A)$
and $H(B)$.]
\end{document}